\documentclass[twocolumn]{aa}
\usepackage{graphicx}
\usepackage{txfonts}
\usepackage{natbib}
\bibpunct{(}{)}{;}{a}{}{,}

\begin{document}

\title{Modeling of iron K lines: radiative and Auger decay data for
Fe~{\sc ii}--Fe~{\sc ix}}

\author{
P. Palmeri\inst{1}\fnmsep\thanks{Research Associate, Department of
               Astronomy, University of Maryland, College Park, MD 20742}
\and
C. Mendoza\inst{1}\fnmsep\thanks{Current address: Centro de F\'{\i}sica,
                   IVIC, Caracas 1020A, Venezuela}
\and
T. R. Kallman \inst{1}
\and
M. A. Bautista \inst{2}
\and
M. Mel\'endez \inst{2}
}
\institute{
NASA Goddard Space Flight Center, Code 662, Greenbelt, MD 20771, USA
\and
Centro de F\'{\i}sica, Instituto Venezolano de Investigaciones
Cient\'{\i}ficas (IVIC), PO Box 21827, Caracas 1020A, Venezuela
}

\offprints{T. R. Kallman, \email{timothy.r.kallman@nasa.gov}}

\date{Received ; }


\abstract{

A detailed analysis of the radiative and Auger de-excitation
channels of K-shell vacancy states in Fe~{\sc ii}--Fe~{\sc ix}
has been carried out. Level energies, wavelengths, $A$-values,
Auger rates and fluorescence yields have been calculated for the
lowest fine-structure levels populated by photoionization of the
ground state of the parent ion. Different branching ratios,
namely K$\alpha_2$/K$\alpha_1$, K$\beta$/K$\alpha$, KLM/KLL, KMM/KLL,
and the total K-shell fluorescence yields, $\omega_K$, obtained in the present
work have been compared with other theoretical data and
solid-state measurements, finding good general agreement with the
latter. The K$\alpha_2$/K$\alpha_1$ ratio is found to be
sensitive to the excitation mechanism. From these comparisons it
has been possible to estimate an accuracy of $\sim$10\% for the
present transition probabilities.

}
\authorrunning{Palmeri et al.}
\titlerunning{K lines in Fe~{\sc ii--ix}}

\maketitle

\section{Introduction}

The iron K lines appear in a relatively
unconfused spectral region and have a high
diagnostic potential. The study of these lines has been encouraged
by the quality spectra emerging from {\em Chandra}
and by the higher resolution expected from
{\em Astro-E} and {\em Constellation-X}.  In addition there is
a shortage of accurate and complete level-to-level atomic
data sets for the K-vacancy states of the Fe isonuclear sequence, in
particular for the lowly ionized species. This
undermines line identification and realistic spectral modeling. We are
currently remedying this situation by systematic calculations using
suites of codes developed in the field of computational
atomic physics. Publicly available packages have been chosen rather than
in-house developments.
In this context, complete data sets for the $n=2$
K-vacancy states of the first row, namely Fe~{\sc xviii}--Fe~{\sc xxv},
have been reported earlier by \citet{bau03} and
\citet{pal03}, to be referred to hereafter as Paper~I and Paper~II, and
for the second row (Fe~{\sc x}--Fe~{\sc xvii}) by \citet{men03}, to be
referred to as Paper~III.

The K lines from Fe species with
electron occupancies $N>17$ have been studied very little.
\citet{jac86} have computed fluorescence probabilities in a
frozen-core approximation for vacancies among the $n\ell$ subshells of
the Fe isonuclear sequence.  \citet{kaa93} have calculated
the inner-shell decay processes for all the ions from Be to Zn by
scaling published Auger and radiative rates in neutrals.
Both of these studies ignore multiplets and fine-structure. Otherwise,
the bulk of the data in the literature is devoted to solid-state
iron. Regarding wavelengths and line intensity ratios, numerous references
are listed in \citet{holz97} who measure the K$\alpha_{1,2}$ and
K$\beta_{1,3}$ emission lines of the 3d transition metals using a
high-precision single-crystal spectrometer.
Fewer publications are available on the experimental K Auger
spectra: \citet{kov87} have derived the KLM/KLL and KMM/KLL ratios from
the Auger electron spectrum of iron produced by $^{57}$Co decay, and
\citet{nem96} have measured the KLL and KLM spectra of the 3d
transition metals but have not determined the KLM/KLL ratio.
Concerning K-shell fluorescence yields, measurements
covering the period 1978--93 have been reviewed by \citet{hubb94}
following major compilations by \citet{bam72} and \citet{kra79}.

The present work is a detailed analysis of the radiative and
Auger de-excitation channels of the K-shell vacancy states in the third-row
species Fe~{\sc ii}--Fe~{\sc ix}. Energy levels, wavelengths, $A$-values,
Auger rates and fluorescence yields have been computed
for the lowest fine-structure levels in configurations obtained by
removing a 1s electron from the ground configuration of the parent ion.
In Section~2 the numerical method is briefly described. Section~3 outlines
the decay trees and selection rules. Results and discussions are given
in Section~4, followed by a summary and conclusions (Section~5).
All the atomic data calculated in this work are available in the
electronic Tables~3--4.


\section{Numerical method}

In Paper~III we conclude that the importance of core-relaxation
effects increases with electron occupancy, so calculations of third-row iron ions
require a computational platform which is well suited to treating these effects.
For this reason the calculations reported here are carried out using the
{\sc hfr} package by \citet{cow81} although {\sc autostructure}
\citep{bad86,bad97} is also heavily used for comparison purposes. In {\sc hfr}
an orbital basis is obtained for each electronic configuration
by solving the Hartree--Fock equations for the spherically
averaged atom. The equations are derived from the application
of the variational principle to the configuration average
energy and include relativistic corrections, namely the
Blume--Watson spin--orbit, mass--velocity and the one-body Darwin terms.
The eigenvalues and eigenstates thus obtained
are used to compute the wavelength and $A$-value for each possible transition.
Autoionization rates are calculated in a perturbation theory scheme
where the radial functions of the initial and
final states are optimized separately, and configuration interaction is
accounted for only in the autoionizing state. Configuration interaction
is taken into account among the following configurations: $({\rm 3d+4s})^M$,
$[{\rm 3p}]({\rm 3d+4s})^{M+1}$, $[{\rm 2p}]({\rm 3d+4s})^{M+1}$
and $[{\rm 1s}]({\rm 3d+4s})^{M+1}$ where $[n\ell]$ stands for a hole in
the $n\ell$ subshell and $(n\ell+n'\ell')^M$ represents all possible
distributions of $M$ electrons among the $n\ell$ and $n'\ell'$ shells,
$M$ ranging from $M=0$ in Fe~{\sc ix} to $M=7$ in Fe~{\sc ii}.

Given the complexity of $[{\rm 2p}][{\rm 3p}]$ double-vacancy
channels in ions with an open 3d shell, the level-to-level computation of the Auger rates
with {\sc hfr} and {\sc autostructure} proved to
be intractable. However, average values can be
used for all the levels taking advantage of the near constancy of
the total Auger widths in ions for which the KLL channels reduce
the outer configuration to a spectator (see Paper~III).
Therefore, we employ the formula given in \citet{pal01} for the
single-configuration average (SCA) Auger decay rate
\begin{equation}
A^{\rm SCA}_a = \frac{\sum_{i} (2J_i+1) A_a(i)}{\sum_{i} (2J_i + 1)}
= \frac{s}{g} A_a( n \ell n' \ell' \rightarrow n'' \ell'' \varepsilon \ell''')
\label{asca1}
\end{equation}
where the sum runs over all the levels of the autoionizing configuration,
$s/g$ is a statistical factor given in Eqs.~(15--16) of \citet{pal01}
that contains the dependence on the active shell
($n\ell$, $n'\ell'$, $n''\ell''$) occupancy, and
$A_a(n \ell n' \ell' \rightarrow n'' \ell'' \varepsilon \ell''')$
is the two-electron autoionization rate which is a function of the radial
integrals and for which the complete expression is given in Eq.~(11)
of \citet{pal01}. The calculated Auger rates are expected be as accurate
as those obtained in a level-to-level single-configuration approach.


\section{Photoionization selection rules and decay trees}

We have focused our calculations on the Fe K-vacancy states populated by
photoionization of the ground state of the parent ion
\begin{equation}
\mu\ \ ^{(2S+1)}L_{J} + \gamma \longrightarrow
                            [{\rm 1s}]\mu\ \ ^{(2S'+1)}L'_{J'} + e^-
\label{kpi}
\end{equation}
where $\mu$ is the outer configuration of the ground state, i.e.
${\rm 3d}^6 {\rm 4s}^2$ in Fe~{\sc i}, ${\rm 3d}^6 {\rm 4s} $ in Fe~{\sc ii}
and ${\rm 3d}^N$ ($N=1-6$) in Fe~{\sc iii}--Fe~{\sc viii}, $[{\rm 1s}]$
denoting a K hole. In Eq.~(\ref{kpi}), K-shell photoionization leads
to a p wave with selection rules specified by \citet{rau76}:
$L' = L$,  $S'-S = \pm 1/2$ and $J'-J = \pm 1/2,\pm 3/2$. Therefore,
only few states are expected to be populated.

The radiative and Auger decay manifolds of a K-vacancy configuration
$[{\rm 1s}]\mu$ can be outlined as follows:
\begin{itemize}
\item{Radiative channels}
\begin{eqnarray}
[{\rm 1s}]\mu & \stackrel{{\rm K}\beta}{\longrightarrow} &
                  [{\rm 3p}]\mu+\gamma_\beta \\
             & \stackrel{{\rm K}\alpha}{\longrightarrow} &
                  [{\rm 2p}]\mu+\gamma_\alpha
\end{eqnarray}
\item{Auger channels}
\begin{eqnarray}
[{\rm 1s}]\mu & \stackrel{\rm KMM(p)}\longrightarrow &
               \left\{
                  \begin{array}{l}
                  \mu^{-2}+e^- \\
                  \left[{\rm 3p}\right]\mu^{-1}+e^- \\
                  \left[{\rm 3s}\right]\mu^{-1}+e^-
               \end{array}
                \right\}\\
              & \stackrel{\rm KMM(s)}\longrightarrow &
               \left\{\begin{array}{l}
                  \left[{\rm 3p}\right]^2\mu+e^- \\
                  \left[{\rm 3s}\right]\left[{\rm 3p}\right]\mu+e^- \\
                  \left[{\rm 3s}\right]^2\mu+e^-
               \end{array}\right\}\\
              & \stackrel{\rm KLM(p)}\longrightarrow &
                  \left\{
                     \begin{array}{l}
                       \left[{\rm 2s}\right]\mu^{-1}+e^- \\
                       \left[{\rm 2p}\right]\mu^{-1}+e^-
                     \end{array}\right\}  \\
              & \stackrel{\rm KLM(s)}\longrightarrow &
                  \left\{
                     \begin{array}{l}
                       \left[{\rm 2s}\right]\left[{\rm 3p}\right]\mu+e^- \\
                       \left[{\rm 2p}\right]\left[{\rm 3p}\right]\mu+e^- \\
                       \left[{\rm 2s}\right]\left[{\rm 3s}\right]\mu+e^- \\
                       \left[{\rm 2p}\right]\left[{\rm 3s}\right]\mu+e^-
                     \end{array}\right\}  \\
              & \stackrel{\rm KLL}\longrightarrow &
                  \left\{
                     \begin{array}{l}
                        \left[{\rm 2s}\right]^2\mu+e^- \\
                        \left[{\rm 2s}\right]\left[{\rm 2p}\right]\mu+e^- \\
                        \left[{\rm 2p}\right]^2\mu+e^-
                     \end{array}\right\}
\end{eqnarray}
\end{itemize}
where the negative exponent of $\mu$ stands for the number of electrons that
have been extracted from the outer subshells.
In the radiative channels, forbidden and two-electron transitions have been
excluded as it has been confirmed by calculation that
they display very small transition probabilities ($\log A_{\rm r} < 11$).
Therefore, the two main photo-decay pathways are the result of the
$2{\rm p}\rightarrow {\rm 1s}$ and
$3{\rm p}\rightarrow {\rm 1s}$ single electron jumps that
give rise respectively
to the K$\alpha$ ($\sim {\lambda}1.94$) and K$\beta$ ($\sim {\lambda}1.75$)
arrays.

It has also been numerically verified that the participator Auger
channels (KMM(p) and KLM(p)) contribute less than one percent to
the total Auger widths, and hence, they have not been taken into
account. Of primary interest in the present work is the branching
ratios of the K$\alpha$, K$\beta$, KMM, KLM and KLL channels and
their variations with electrons occupancy $N$.

\section{Results and discussion}

\begin{figure}
\centering
\resizebox{\hsize}{!}{\includegraphics{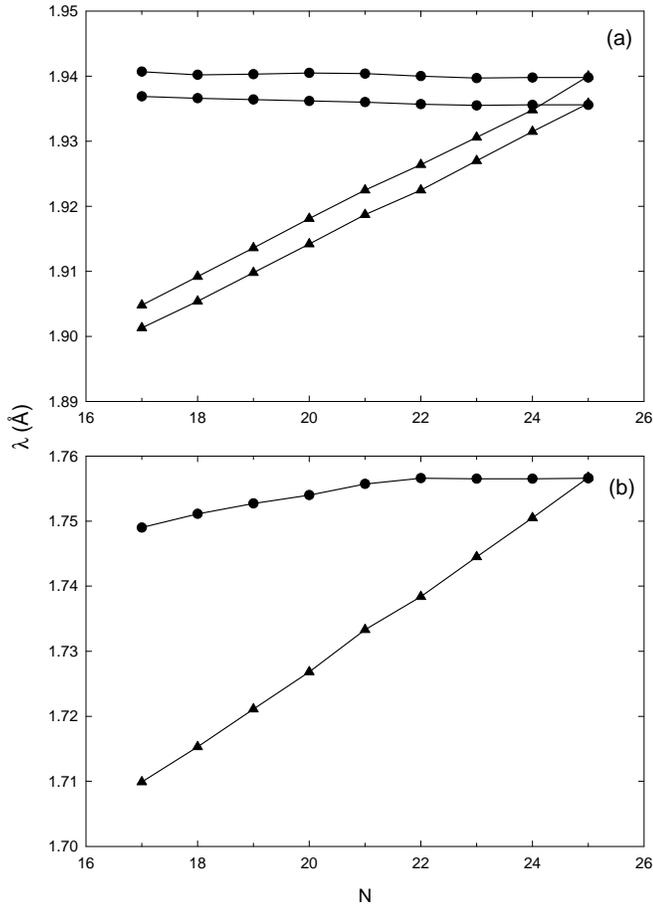}}
\caption{Comparison of centroid wavelengths for (a)
K$\alpha$ and (b) K$\beta$ UTAs in Fe ions with $17\leq N\leq 25$.
Filled circles: this work. Filled triangles: \citet{kaa93}.}
\label{fig1}
\end{figure}
\begin{table}
\centering
\caption[]{Comparison of centroid wavelengths (\AA) for the K$\alpha$ and
K$\beta$ unresolved transition arrays in Fe ions with
$17\leq N\leq 25$. The {\sc hfr} wavelengths have been
weighted with the fluorescence yields. Experimental wavelengths for
Fe~{\sc x} ($N=17$) are from \citet{dec95} and for solid-state iron
from \citet{holz97}.}
\label{table1}
\begin{tabular}{llllll}
\hline\hline
&\multicolumn{2}{c}{HFR} &&\multicolumn{2}{c}{Expt}\\
\cline{2-3}\cline{5-6}
$N$ & K$\alpha$ & K$\beta$     && K$\alpha$   & K$\beta$ \\
\hline
 17 & 1.9369 & 1.7488 && 1.9388(5)   &            \\
    & 1.9407 & 1.7493 && 1.9413(5)   &            \\
 18 & 1.9366 & 1.7511 &&             &            \\
    & 1.9402 &        &&             &            \\
 19 & 1.9364 & 1.7527 &&             &            \\
    & 1.9403 &        &&             &            \\
 20 & 1.9362 & 1.7540 &&             &            \\
    & 1.9405 &        &&             &            \\
 21 & 1.9360 & 1.7557 &&             &            \\
    & 1.9404 &        &&             &            \\
 22 & 1.9357 & 1.7566 &&             &            \\
    & 1.9400 &        &&             &            \\
 23 & 1.9355 & 1.7565 &&             &            \\
    & 1.9397 &        &&             &            \\
 24 & 1.9356 & 1.7565 &&             &            \\
    & 1.9398 &        &&             &            \\
 25 & 1.9356 & 1.7566 && 1.936041(3) & 1.756604(4)\\
    & 1.9398 &        && 1.939973(3) &            \\
\hline
\end{tabular}
\end{table}

Centroid wavelengths for the K$\alpha$ and K$\beta$ unresolved
transition arrays (UTAs) computed with {\sc hfr} are presented in
Table~\ref{table1}, including also a comparison with experiment.
For the K$\alpha_1$ and K$\alpha_2$ lines, it can be seen that
the agreement with the solid-state measurements by \citet{holz97}
($\sim 0.5$ m\AA) is somewhat better than that with the EBIT
results for Fe~{\sc x} \citep{dec95} of within 2 m\AA, and the
slight blueshift with increasing $N$ predicted by {\sc hfr} is
consistent with experiment. A small redshift with $N$ is also
found for the K$\beta$ array. On the other hand, as shown in
Figure~\ref{fig1}, the present findings contrast with the steeper
redshift for both UTAs obtained from the data by \citet{kaa93}.

\begin{figure}
\centering
\scalebox{.85}{\resizebox{\hsize}{!}{\includegraphics{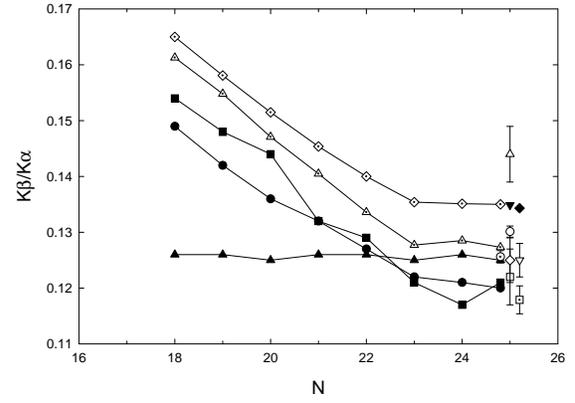}}}
\caption{Comparison of K$\beta$/K$\alpha$ intensity ratios for Fe ions
with $18\leq N\leq 25$. Filled circles: {\sc hfr}, this work. Dotted
upright triangles: {\sc autostructure}, this work. Dotted diamonds:
{\sc mcdf-sal}, this work. Filled upright triangles: \citet{kaa93}.
Filled squares: \citet{jac86}. Filled diamonds: \citet{sco74}. Filled
inverted triangles: \citet{jan89}. Circles: \cite{per87}. Upright
triangles: \citet{holz97}. Squares: \citet{rao86}. Diamonds: \citet{ber78}.
Inverted triangles: \citet{sal72}. Dotted circles: \citet{sli72}. Dotted
squares: \citet{han70}.}
\label{fig2}
\end{figure}

\begin{table}
\centering
\caption[]{Comparison of {\sc hfr} K$\alpha_{2}$/K$\alpha_{1}$ ratios
for Fe ions ($18\leq N\leq 25$) with experiment and previous theoretical
estimates.}
\label{table2}
\begin{tabular}{llllll}
\hline\hline
$N$ & Ion & HFR1$^{\mathrm{a}}$ & HFR2 $^{\mathrm{b}}$ &
Expt.  & Theory \\
\hline
18 & Fe~{\sc ix}   & 0.7000 & 0.5912 &          &  \\
19 & Fe~{\sc viii} & 0.5780 & 0.6015 &          &  \\
20 & Fe~{\sc vii}  & 0.4541 & 0.5821 &          &  \\
21 & Fe~{\sc vi}   & 0.5545 & 0.5538 &          &  \\
22 & Fe~{\sc v}    & 0.5852 & 0.5346 &          &  \\
23 & Fe~{\sc iv}   & 0.4426 & 0.5143 &          &  \\
24 & Fe~{\sc iii}  & 0.4438 & 0.5032 &          &  \\
25 & Fe~{\sc ii}   & 0.4476 & 0.4969 & 0.51(2)$^{\mathrm{c}}$   & 0.5107$^{\mathrm{g}}$ \\
   &               &        &        & 0.507(10)$^{\mathrm{d}}$ & \\
   &               &        &        & 0.506$^{\mathrm{e}}$     & \\
   &               &        &        & 0.4998$^{\mathrm{f}}$    & \\
\hline
\end{tabular}
\begin{list}{}{}
\item[$^{\mathrm{a}}$] Includes only the decay lines from the
K-vacancy levels listed in Table~3\\
\item[$^{\mathrm{b}}$] Includes all the decay lines from the
$[{\rm 1s}]({\rm 3d+4s})^{M+1}$ complex\\
\item[$^{\mathrm{c}}$] \citet{holz97}\\
\item[$^{\mathrm{d}}$] \citet{mcg71}\\
\item[$^{\mathrm{e}}$] \citet{sal70}\\
\item[$^{\mathrm{f}}$] \citet{wil33}\\
\item[$^{\mathrm{g}}$] \citet{sco74}\\
\end{list}
\end{table}

In Table~\ref{table2}, K$\alpha_2$/K$\alpha_1$ intensity ratios are
tabulated for different ionization stages. Two cases have been considered:
HFR1, only the decay lines from the K-vacancy levels populated by the
photoionization of the respective ground state are included; and HFR2, all the
transitions from the levels belonging to the $[{\rm 1s}]({\rm 3d+4s})^{M+1}$
complex are taken into account. Noticeable differences in the ratios from
these two cases may be appreciated indicating sensitivity to the excitation
mechanism. The HFR2 ratio for Fe~{\sc ii} is closer to the Dirac--Fock
value of \citet{sco74}, who also considered all the K-vacancy states, and
to the solid-state experiments \citep{holz97,mcg71,sal70,wil33}.

\begin{figure}
\centering
\scalebox{.85}{\resizebox{\hsize}{!}{\includegraphics{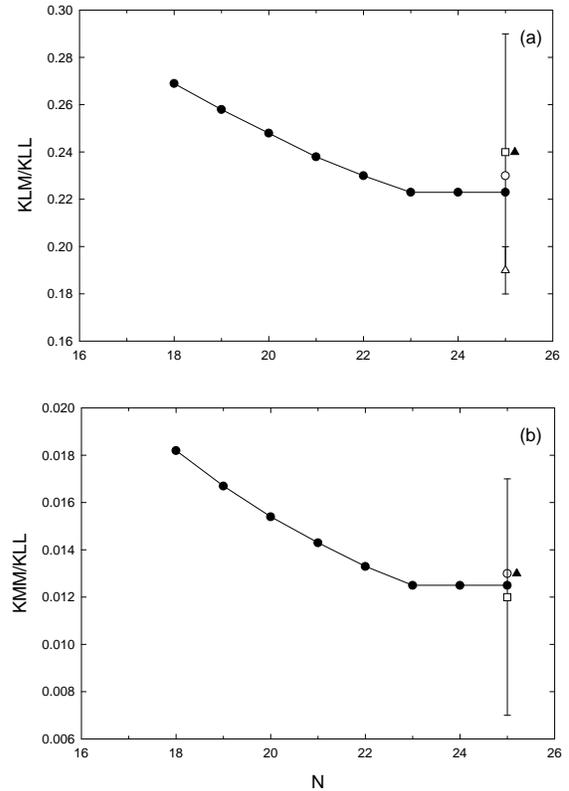}}}
\caption{Comparison of Auger rates for Fe ions with $18\leq N\leq 25$:
(a) KLM/KLL and (b) KMM/KLL ratios. Filled circles:
{\sc hfr}, this work. Upright filled triangles: \citet{che79}. Circles:
\citet{bha70}. Squares: \citet{kov87}. Upright triangles: \citet{meh63}.}
\label{fig3}
\end{figure}

In Figure~\ref{fig2}, theoretical and experimental K$\beta$/K$\alpha$ intensity
ratios are plotted as a function of electron number. Most experimental ratios
\citep{rao86,ber78,sal72,sli72,han70} have been scaled down by 8.8\%
in order to extract the radiative-Auger (5\%) and K$\beta_5$ (3.8\%) satellite
contributions from the K$\beta$ UTA which theory does not include
\citep{ver00}. The value quoted in \citet{per87} does not take into
account the blend with the K$\beta_5$ satellite and has been corrected
accordingly. With the exception of the values by \citet{kaa93}, theory
predicts a decrease of the ratio with $N$, and the theoretical scatter
is comparable with that among the solid-state experiments of just under 20\%.
The computed results by {\sc hfr}, {\sc autostructure} and \citet{jac86}
for $N=25$ are in good agreement with the bulk of the experimental values
\citep{per87,rao86,ber78,sal72,sli72,han70}. \citet{holz97} mention a
possible systematic deviation in one of their corrections
to explain the discrepancy of their data with other measurements.
On the other hand, the Dirac--Fock ratios by \citet{sco74} and \citet{jan89}
at $N=25$ appear significantly higher. We have therefore proceeded to
verify these values by using the same code ({\sc mcdf-sal}) as in \citet{jan89},
and as shown in Figure~\ref{fig2}, they are accurately reproduced; moreover,
a similar decrease in the ratio with $N$ is also obtained with this method.
The spread of the different data sets in this comparison indicates a probable
accuracy of our {\sc hfr} transition probabilities of $\sim 10\%$.

\begin{figure}
\centering
\scalebox{.85}{\resizebox{\hsize}{!}{\includegraphics{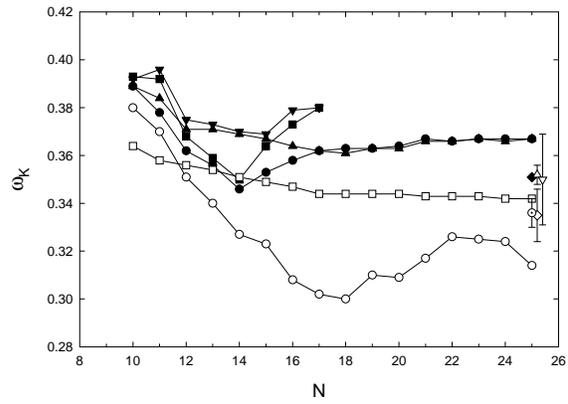}}}
\caption{Comparison of K-shell fluorescence yields, $\omega_K$, for Fe ions
$10\leq N\leq 25$. Filled circles: {\sc hfr} value obtained for the first
K-vacancy level, $10\leq N\leq 17$ (Paper~III), $18\leq N\leq 25$ (this work).
Filled squares: {\sc autostructure} value obtained for the first K-vacancy level, Paper~III.
Filled upright triangle: {\sc hfr} value obtained by averaging on all the K-vacancy levels,
$10\leq N\leq 17$ (Paper~III), $18\leq N\leq 25$ (this work).
Filled inverted triangle: {\sc autostructure} value obtained by averaging
on all the K-vacancy levels, Paper~III.
Circles: \citet{jac86}. Squares: \citet{kaa93}. Filled diamonds:
value recommended by \citet{hubb94}. Upright triangle: \citet{sol92}. Inverted triangle:
\citet{pio92}. Diamond: \citet{bha81}. Dotted circle: \citet{sin90}.}
\label{fig4}
\end{figure}

\begin{figure*}
\centering
\resizebox{\hsize}{!}{\includegraphics[width=12cm]{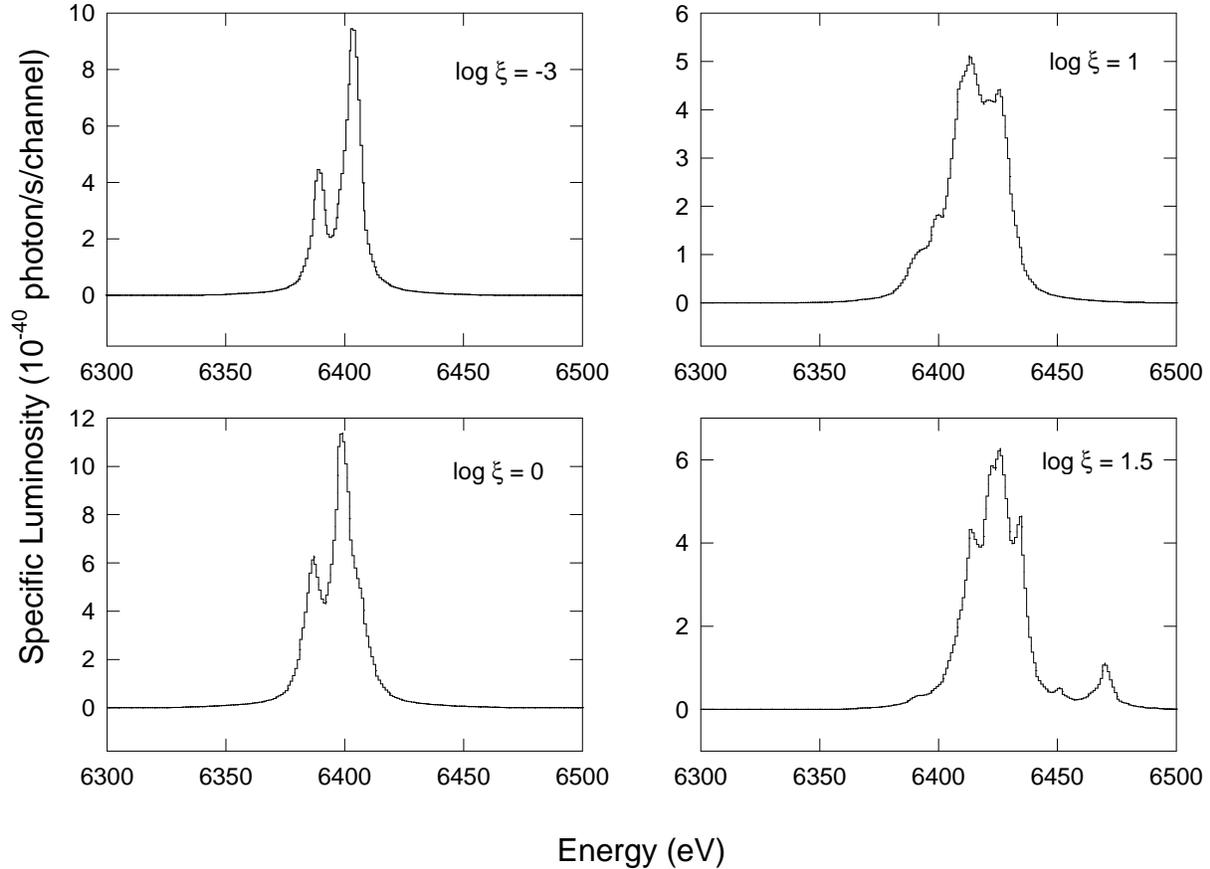}}
\caption{{Simulations carried out near 6.4 keV with the \sc xstar} modeling
code for several values of the ionization parameter $\xi$.}
\label{fig5}
\end{figure*}

A decrease with $N$ is also predicted by {\sc hfr} for the Auger KLM/KLL and
KMM/KLL ratios (see Figure~\ref{fig3}). For Fe~{\sc ii}, present results
are in excellent accord with the measurements of \citet{kov87} and other
theory \citep{che79,bha70} but somewhat higher than the older experimental
estimate of \citet{meh63}. The total K-shell fluorescence yields, $\omega_K$,
for Fe ions with $10\leq N\leq 25$ are presented in Figure~\ref{fig4}.
In both {\sc hfr} and {\sc autostructure} data sets, the fluorescence yields have
been computed for fine-structure K-vacancy levels:
\begin{equation}
\omega_K(i)~=~\frac{A_r(i)}{A_r(i)+A_a(i)} \ .
\label{yield}
\end{equation}
In order to display the effect of the level population, we report in
Figure~\ref{fig4}
the value for the first K-vacancy level and the yield averaged over
all the K-vacancy levels.
As expected, the fluorescence yield is independent of the population mechanism
in the third row ions where the K$\beta$, KLM and KMM decay channels become
spectators. Given the complexity of level-to-level calculations of Auger rates,
it has not been possible to compute fluorescence yields with
{\sc autostructure} for $N>17$. The most recent experimental measurements in the solid
\citep{sol92,pio92,sin90,bha81} are plotted along with the value recommended by \citet{hubb94}.
They show a scatter within 10\% and are slightly lower than our {\sc hfr} value
for $N=5$, i.e. 5 to 10\% with respect to \citet{hubb94} and \citet{bha81} respectively.
This is not a significant discord considering the expected accuracy of our fluorescence yields ($\sim 20\%$)
as discussed in Paper~III.
As mentioned in the previous section, the missing participator Auger channels will not contribute
significantly to the Auger widths (less than 1\%). Concerning the shake-up channels, they will not
affect the Auger widths because they are due to mixing between valence shell configurations which
have the same Auger widths. We have verified that this is the case in a simple
calculation ($[{\rm 1s}]({\rm 3d + 4s})^8 \rightarrow [{\rm 2p}]^2 ({\rm 3d + 4s})^8 + k{\rm s}$).
Data by \citet{kaa93} have been corrected so as to reproduce the best solid-state
yields compiled in \citet{bam72} and are therefore somewhat lower than our values.
Results by \citet{jac86} for $N>12$ predict a substantially lower yield.

In order to evaluate the impact of the new atomic data on Fe K-line
modeling, the data sets generated in the present work and those in
Papers~I--III have been included in the {\sc xstar} modeling code
\citep{kal01}. Runs for different ionization parameters, $\xi$, are shown in
Figure~\ref{fig5}. The other parameters have been assigned the following values:
cosmic abundances; a gas column density of $10^{16}$ cm$^{-2}$;
a gas density of $10^{12}$ cm$^{-3}$; an X-ray source luminosity of 10$^{38}$ erg/s;
a power-law index of 1; and energy bins (or channel widths) of 1 eV.
In Figure~\ref{fig5}, one clearly sees that the UTA centroid
is redshifted when $\xi$ goes from 0.001 to 1 and then blueshifted for $\xi > 1$.
The shape of the UTA changes also considerably for $\xi > 1$ where the ionization
balance favors the first and second row ions.

Electronic Tables~3--4 list radiative and Auger widths for 295 energy
levels; and wavelengths, $A$-values and fluorescence yields for 396 transitions.


\section{Summary and conclusions}

Following the findings of our previous study on the second-row iron ions
(Paper~III), the {\sc hfr} package \citep{cow81} has been selected to compute
level energies, wavelengths, decay rates and fluorescence yields for the
K-vacancy states in Fe~{\sc ii}--Fe~{\sc ix}. The calculations have focused
states populated by ground-state photoionization. Due to the
complexity of the level-to-level Auger calculations, we have employed a compact
formula \citep{pal01} to compute Auger widths from the {\sc hfr} radial
integrals.

The {\sc hfr} centroid wavelengths for the K$\alpha$ and K$\beta$ UTAs
in Fe~{\sc ii} reproduce the solid-state measurements \citep{holz97} to
better than 1 m\AA. Moreover, the red shift predicted by {\sc hfr} for
the K$\alpha$ lines in species with higher ionization stage is in
accord with the EBIT wavelengths in Fe~{\sc x} \citep{dec95} thus
contradicting the previous trend specified by \citet{kaa93}.
We have also carried out extensive comparisons of different {\sc hfr}
branching ratios, namely K$\alpha_2$/K$\alpha_1$, K$\beta$/K$\alpha$,
KLM/KLL, KMM/KLL and $\omega_K$, with other theoretical and experimental
data. The present ratios for Fe~{\sc ii} are in good agreement with the
solid-state measurements, and the K$\alpha_2$/K$\alpha_1$ ratio has been
found to be sensitive to the excitation mechanism. It has been possible
from these comparisons to estimate an accuracy of $\sim$10\% for the
{\sc hfr} transition probabilities.

The new atomic data sets for the whole Fe isonuclear sequence
that have emerged from the present project are providing a more
reliable platform for the modeling of Fe K lines. Preliminary
simulations of the emissivity of a photoionized gas with {\sc
xstar} \citep{kal01} have shown K$\alpha$ line profiles and
wavelength shifts that are sensitive to the ionization level of
the gas. Further work will therefore be concerned with improving
diagnostic capabilities by means of accurate K-shell
photoionization and electron impact excitation cross sections.


\begin{acknowledgements}
PP acknowledges a Research Associateship from University of
Maryland and CM a Senior Research Associateship from the National Research
Council. This work is partially funded by FONACIT, Venezuela,
under contract No. S1-20011000912.
\end{acknowledgements}



\end{document}